\documentclass[twocolumn,english, aip,amsmath,amssymb,reprint]{revtex4-1}
\usepackage{float}
\usepackage{amsmath}
\makeatletter
\usepackage{graphicx}
\usepackage{bm}
\usepackage[utf8]{inputenc}
\usepackage[T1]{fontenc}
\usepackage{mathptmx}

\makeatother

\usepackage{babel}
\begin{document}

\title{Algorithms and Methods for High Precision Computer Simulations of
Cyclotrons for Proton Therapy: The Case of SC202}

\author{Taisia Karamysheva}
\email{taisia.karamysheva@gmail.com}

\affiliation{Federal Research Center ``Computer Science and Control'' of the
Russian Academy of Sciences, Moscow, Russia}

\affiliation{Joint Institute for Nuclear Research, Dubna, Russia }

\author{Oleg Karamyshev}

\author{Vladimir Malinin}

\author{Dmitry Popov}

\affiliation{Joint Institute for Nuclear Research, Dubna, Russia }

\date{\today}
\begin{abstract}
Effective and accurate computer simulations are highly important in
accelerator design and production. The most difficult and important
task in cyclotron development is magnetic field simulations. It is
necessary to achieve the accuracy of the model that is higher than
the tolerance for the magnetic field in the real magnet. An accurate
model of the magnet and other systems of the cyclotron allows us to
perform beam tracking through the whole accelerator from the ion source
to extraction. While high accuracy is necessary in the late stages
of research and development, high performance of the simulations and
ability to swiftly analyze and apply changes to the project play a
key role in the early stages of the project. Techniques and algorithms
for high accuracy and performance of the magnet simulations have been
created and used for the development of the SC202 cyclotron for proton
therapy, which is under production by collaboration between JINR (Dubna,
Russia) and ASIPP (Hefei, China). 
\end{abstract}
\maketitle

\section{Introduction}

Two cyclotrons are planned to be manufactured in China: one will operate
in Hefei cyclotron medical center, the other will replace Phasotron
in Medico-technical complex JINR Dubna \cite{RefB,RefJ} and will
be used to treat cancer with protons. We developed computer programs
for cyclotron design and implemented them in MATLAB/Simulink environment
\cite{Matlab}. MATLAB software is widely used in accelerator physics
\cite{Wolski,AT,Campo}. There are number of MATLAB-based application
programs for particle tracking. The \textquotedbl{}Simple Accelerator
Modelling in Matlab\textquotedbl{} (SAMM) \cite{Wolski} code is a
set of Matlab routines for modelling beam dynamics in high energy
particle accelerators. Accelerator Toolbox (AT) is a collection of
tools to model storage rings \cite{AT}. We use MATLAB to create input
data sets and to access them via Simulink block diagrams during simulation.
We perform thousands of particle tracking simulations in the cyclotron
in parallel and analyze and visualize the data in MATLAB.

\section{Magnet design and beam dynamics simulations}

3D magnet simulation is the most important part when it comes to the
development of an isochronous superconducting cyclotron as it defines
particle motion in the accelerator. Magnet modeling goes hand in hand
with beam dynamics modeling. 

Beam dynamics simulations solve the system of differential equations
of particle motion with field maps from FEM (finite element method)
codes or field measurements with different initial coordinates and
velocities. The fourth-order Runge-Kutta method is usually used to
solve the system. Systems of differential equations of motion of charged
particles in an electromagnetic field are derived from the Newton-Lorentz
equations:

\[
\frac{d\vec{V}}{dt}=\frac{q}{m}\sqrt{1-\frac{V^{2}}{c^{2}}}\left\{ \vec{E}+\left[\vec{V}\vec{B}\right]-\frac{1}{c^{2}}\vec{V}(\vec{V}\vec{E})\right\} ,
\]

where $\vec{V}$ is the velocity of a charged particle, $\vec{E}$
and $\vec{B}$ are the electric and magnetic fields correspondigly,

\[
\begin{cases}
r^{\prime\prime}-\dfrac{2r^{\prime2}}{r}-r=\dfrac{q\sqrt{r^{2}+r^{\prime2}+z^{\prime2}}}{mcV}\bigg(rB_{z}-z^{\prime}B_{\varphi}-\\
-\dfrac{r^{\prime}z^{\prime}}{r}B_{r}+\dfrac{r^{\prime2}}{r}B_{z}\bigg)+\dfrac{q}{m}\dfrac{r^{2}+r^{\prime2}+z^{\prime2}}{V^{2}}\left(E_{r}-\dfrac{r^{\prime}}{r}E_{\varphi}\right),\\
z^{\prime\prime}-\dfrac{2r^{\prime}z^{\prime}}{r}=\dfrac{q}{mc}\dfrac{\sqrt{r^{2}+r^{\prime2}+z^{\prime2}}}{V}\bigg(r^{\prime}B_{\varphi}-rB_{r}-\\
-\dfrac{z^{\prime2}}{r}B_{r}+\dfrac{r^{\prime}z^{\prime}}{r}B_{z}\bigg)+\dfrac{q}{m}\dfrac{r^{2}+r^{\prime2}+z^{\prime2}}{V^{2}}\left(E_{z}-\dfrac{z^{\prime}}{r}E_{\varphi}\right),\\
t^{\prime}=\dfrac{1}{c\beta}(r^{2}+r^{\prime2}+z^{\prime2})^{\frac{1}{2}}=\dfrac{1}{V}(r^{2}+r^{\prime2}+z^{\prime2})^{\frac{1}{2}}.
\end{cases}
\]

\begin{figure}
\includegraphics[width=7cm]{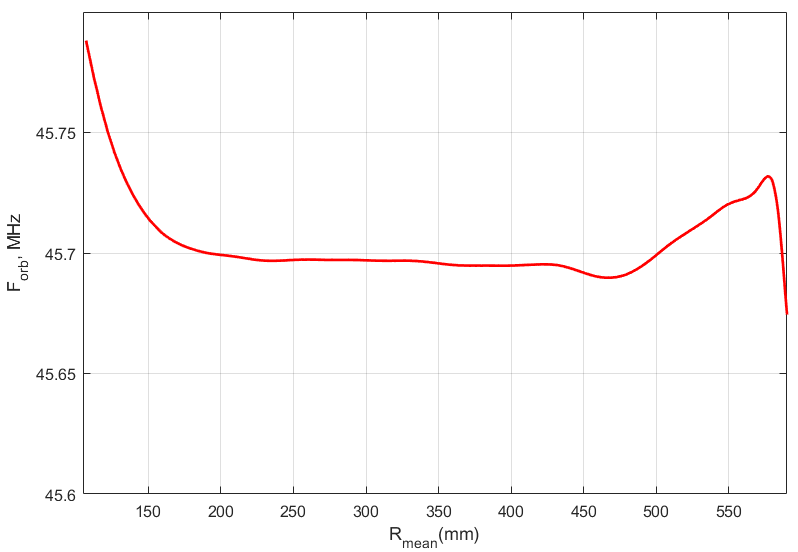}

\caption{\label{fig:Orbital}Orbital frequency of the equlibrium particle along
the radius}

\end{figure}

\begin{figure}
\includegraphics[width=7cm]{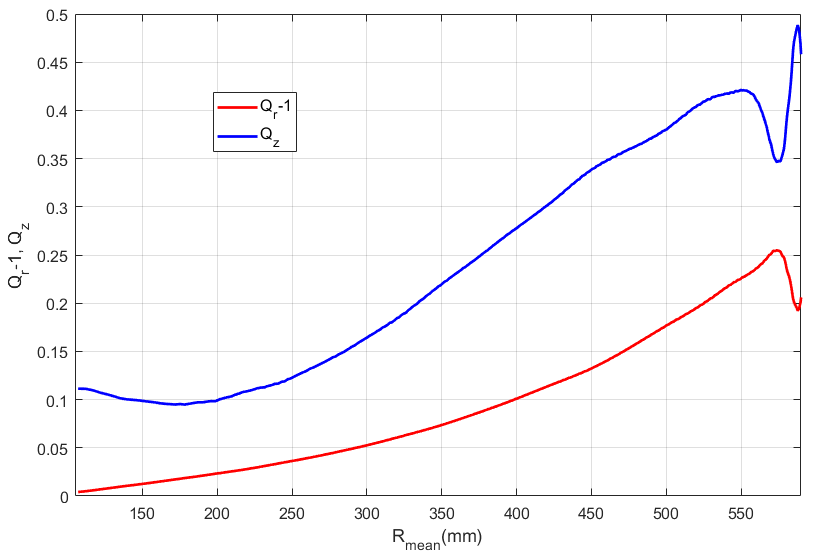}\caption{Betatron frequency of the equlibrium particle along the radius\label{fig:Betatron}}

\end{figure}

In cyclotrons, it is most convenient to use the cylindrical coordinate
system $(r,\varphi,z)$. So the representation of the equations of
motion with an independent variable azimuth angle is widely used in
the design of cyclotrons. Simulink block diagrams for tracking particles
in electromagnetic fields for these equations of motion were created
and used for different cyclotrons \cite{Jongen,Pepan}. Simulink has
a disadvantage: the complexity of the block diagram used to simulate
the particle motion through the whole accelerator can increase drastically
with the number of accelerator systems (harmonic coils, extraction
elements) and beam effects like vacuum losses, space charge effects
etc. Therefore, we rewrote Simulink block diagrams used for cyclotron
design, including the CYCLOPS-like code, in Matlab code. CYCLOPS program
\cite{Gor1} is one of the most reliable tools for dynamics analysis
in cyclotrons. We implemented a version of CYCLOPS using MATLAB R2018a.
The program has an equilibrium orbit search algorithm and performs
calculations of orbital (Fig. \ref{fig:Orbital}) and betatron frequencies
(Fig. \ref{fig:Betatron}). The calculation results can be used for
the isochronisation of the magnetic field map. Another possible way
to use this program is to take its results as input parameters for
3D beam dynamics simulations. The equilibrium orbit search algorithm
requires a 2D magnetic field map (in polar coordinate system) of the
cyclotron median plane as an input parameter. The field map is preprocessed
and smoothed as described in section 3. To improve performance, the
map is accessed via Matlab's griddedInterpolant class. The equations
of motion are solved via an ODE solver (ode45 of MATLAB R2018a) employing
the fourth-order Runge-Kutta algorithm. Our implementation allows
simultaneous calculation of a large number of particles, which improves
performance due to lower number of loops and more efficient use of
RAM memory. For 10\textsuperscript{4}particles the calculation time
was 332 seconds (the calculation was executed in a CPU Intel Core
i7-7700k 4,2GHz with 64GB of RAM memory). To achieve necessary accuracy,
without decrease in performance, we used a timestep of $2\pi/720$
s in units of $\omega_{0}^{-1}$ ($\omega_{0}$ is the cyclotron frequency),
so that each turn corresponds to approximately 720 timesteps.

\section{Representation of the electromagnetic field}

When calculating particle trajectories, an electromagnetic field can
be specified analytically, in the form of a field map obtained as
a result of measurements (2D maps) or as a result of model calculations
(3D field maps). The representation in the form of a 3D field map
is obviously the most acceptable for calculations of beam dynamics,
but maps can be huge when high accuracy is needed.

The major problem of R\&D phase of the SC202 project was that 3D simulations
of the magnet were very time-consuming and it was difficult to achieve
acceptable accuracy of the simulated field. In order to fix this issue,
CST studio macros and quick analysis of the magnet field map in MATLAB
have been developed. Scripts for producing 3D magnetic field maps
in CST studio and for reading them into the Matlab workspace were
written. This technique allowed us to increase the number of simulated
magnet geometries to over 5000 in half a year. Such fine tuning allowed
us to optimize the magnet in a very short time.

Using a 2D field map in the median plane for beam dynamics simulations
is a traditional method which is widely used in cyclotron physics.
Vertical and other components of the magnetic field at the distance
$z$ from the median plain can be calculated by: 

\begin{multline*}
B_{r}=z\frac{\partial B}{\partial r}+\frac{\partial}{\partial r}\Bigg[\sum_{n=1}^{\infty}(-1)^{n}\frac{z^{2n+1}}{(2n+1)!}\bigg(\frac{\partial^{2}}{\partial r^{2}}+\\
+\frac{\partial^{2}}{\partial r^{2}}+\frac{1}{r}\frac{\partial}{\partial r}+\frac{1}{r^{2}}\frac{\partial^{2}}{\partial\varphi^{2}}\bigg)^{n}B\Bigg]
\end{multline*}

\begin{multline*}
rB_{\varphi}=z\frac{\partial B}{\partial\varphi}+\frac{1}{r}\frac{\partial}{\partial\varphi}\Bigg[\sum_{n=1}^{\infty}(-1)^{n}\frac{z^{2n+1}}{(2n+1)!}\bigg(\frac{\partial^{2}}{\partial r^{2}}+\frac{1}{r}\frac{\partial}{\partial r}+,\\
+\frac{1}{r^{2}}\frac{\partial^{2}}{\partial\varphi^{2}}\bigg)^{n}B\Bigg]
\end{multline*}

\begin{multline*}
B_{z}=B+\Bigg[\sum_{n=1}^{\infty}(-1)^{n}\frac{z^{2n}}{(2n)!}\bigg(\frac{\partial^{2}}{\partial r^{2}}+\frac{1}{r}\frac{\partial}{\partial r}+\\
+\frac{1}{r^{2}}\frac{\partial^{2}}{\partial\varphi^{2}}\bigg)^{n}B\Bigg]
\end{multline*}

\begin{figure*}[!t]
\includegraphics[width=18cm]{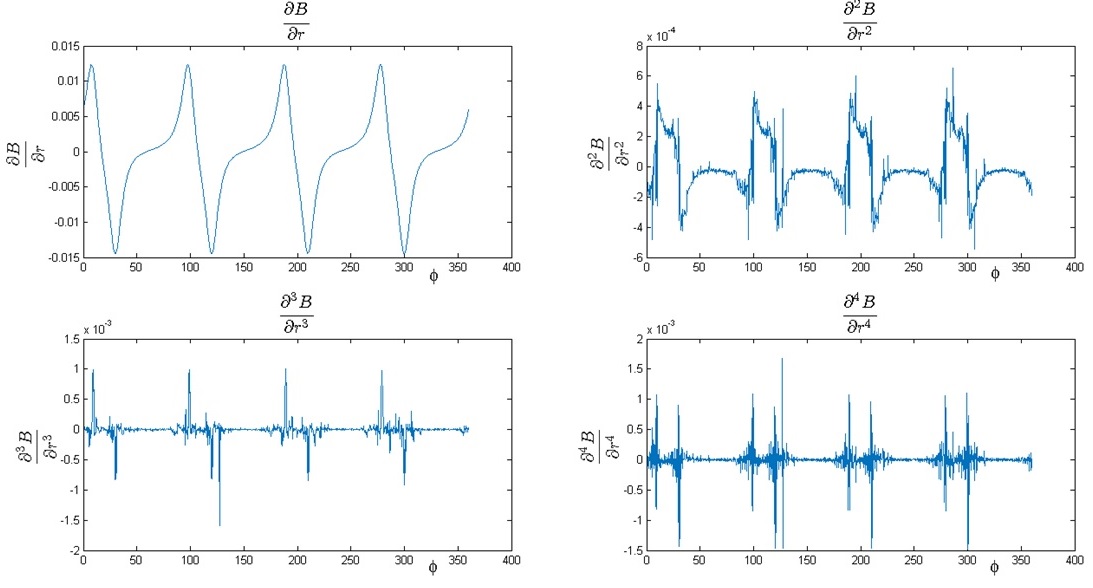}\caption{\label{fig:Deriv1}Derivatives of the magnetic field in the median
plane taken without smoothing algorithms}
\end{figure*}

\begin{figure*}[!t]
\includegraphics[width=13cm]{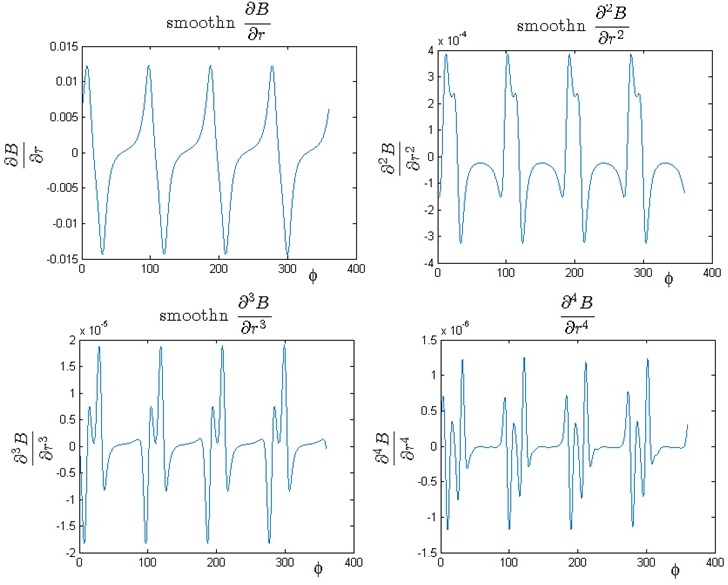}\caption{\label{fig:Deriv2}Derivatives of the magnetic field in the median
plane taken with smoothing algorithms}
\end{figure*}

The expansions are presented for a cylindrical coordinate system,
which is most suitable for cyclotrons. A rigorous proof of these expansions
is presented in \cite{RefP}. The paper reviews the expansion to arbitrary
order. We used the expansion to 2nd order for heavy-ion beam tracking
in the CYTRACK cyclotron, which has been working on track membrane
production in Dubna since 2004 \cite{CYTRACK}, however, when developing
a superconducting cyclotron for medium energy, such as SC202, it is
necessary to study the effects of various nonlinear resonances, therefore
we have added an expansion to 4th order in our codes.

It is important to be able to calculate all derivatives for correct
beam dynamics simulations during research and development, and especially
on the commissioning stage, when real magnetic field will be measured
only on the median plane (Fig.\,\ref{fig:Deriv1}).

The biggest problem of these calculations is that the derivatives
must be taken using the measured or calculated field map, which already
contains errors. Special smoothing algorithms were applied in order
to obtain smooth and realistic derivatives of the magnetic field on
the median plane. In particular, we used a MATLAB function SMOOTHN,
that provides a fast, unsupervised and robust discretized spline smoother
for data of arbitrary dimension \cite{Gar1,Gar2}.

Fig.\,\ref{fig:Deriv2} represents the results, that were obtained
by combining fitting of the field map with spline surface together
with smoothing algorithms.

Due to the symmetrical design of the magnet of the cyclotron (usually
four fold symmetry), the magnetic field in the median plane is a periodical
function from the azimuth angle. As almost every periodical function
it could be represented as a Fourier series. Fourier series are easy
to differentiate so all angle derivatives (which is half of all) do
not need smoothing mentioned above.

So, we have the following ways to get a 3D map of the magnetic field: 
\begin{itemize}
\item Straightforward one: we obtain raw 3D field map from the simulation. 
\item Using Maxwell's equations on 2D map in median plane (usually measured)
and some interpolation with smoothing. 
\item Same as previous, but this time we apply Fourier analysis. 
\end{itemize}
These approaches were coded in MATLAB and performed on three packages:
CST, TOSCA and COMSOL. The developed software creates 3D field maps
which can be used for beam dynamic simulations.

\section{Accelerating RF system design}

\begin{figure}[H]
\centering{}\includegraphics[width=7cm]{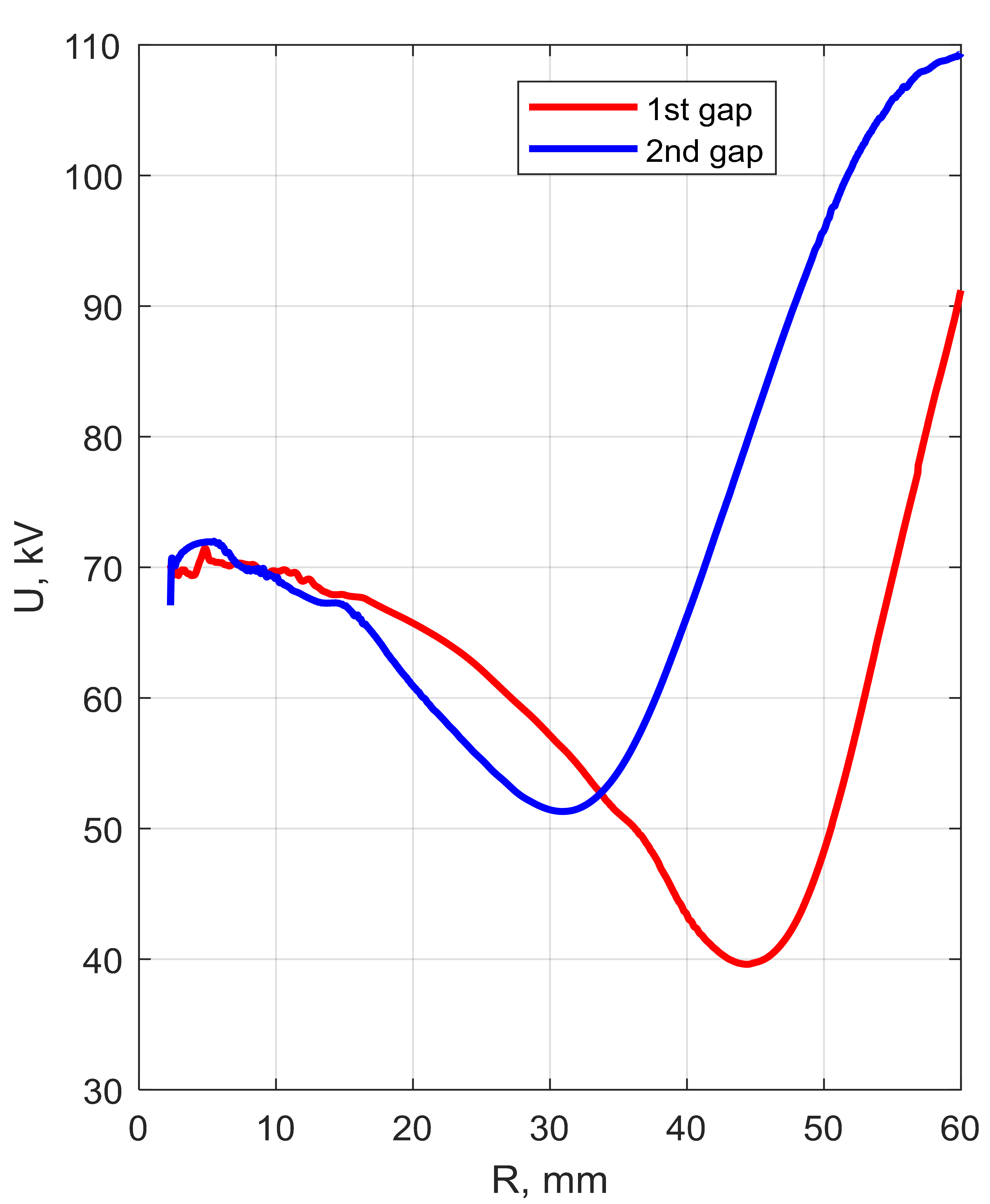}\caption{\label{fig:Voltage}Voltage distribution along the radius}
\end{figure}

It is also important for beam dynamics simulations to use an adequate
representation of the accelerating field. We simulated the accelerating
RF cavity in CST Studio. Both electric and magnetic fields of the
RF cavity from 3D simulations were used in particle tracking from
the ion source to the extraction point of the SC202. Scripts for producing
3D field maps in CST Studio and for reading them into the MATLAB workspace
were written. The code for accelerating voltage distribution calculation
was also created. The voltage value was obtained by integrating the
electric field in the median plane of the resonant cavity (Fig.\,\ref{fig:Voltage}).

\section{Conclusion}

Codes for beam dynamics simulations were improved with new algorithms
where the magnetic field components outside the median plane are calculated
up to fourth order.

Scripts were written for producing 3D magnetic field maps, 3D electric
and magnetic field maps from RF cavity simulations in CST Studio and
for reading them into the MATLAB workspace. MATLAB, being a matrix-oriented
tool, allows us to implement an equilibrium orbit search algorithm
for a large number of particles with different energies with a good
calculation speed.

High accuracy and high efficiency of simulations will help on the
commissioning stage during the shimming of the magnet.

\end{document}